\documentclass[preprint,12pt,number]{elsarticle}

\usepackage{amssymb}
\usepackage{amsmath}
\usepackage{amsthm}
\usepackage{graphicx}
\usepackage{xcolor}
\usepackage{listings}
\usepackage[frozencache,cachedir=mintedcache]{minted}
\usepackage{tcolorbox}
\usepackage{etoolbox}
\usepackage{microtype}
\usepackage{algorithmic}
\usepackage{algorithm}
\usepackage{url}
\usepackage{xurl}
\usepackage[hidelinks]{hyperref}
\usepackage{booktabs}
\usepackage{soul} 
\usepackage{enumitem}
\usepackage[inkscapelatex=false]{svg}
\usepackage{orcidlink}

\setminted{xleftmargin=2em}

\makeatletter
\let\@float@c@listing\@caption
\makeatother

\usepackage{geometry}
\usepackage{textgreek}

\journal{Journal of Computer Languages}

\begin{document}

\begin{frontmatter}

\makeatletter
\patchcmd{\printFirstPageNotes}
  {\ifx\@tnotes\@empty\else\@tnotes\fi}
  {%
    \let\thefootnote\relax\footnotetext{%
      \hspace*{-1.8em}%
      \begin{minipage}[b]{\columnwidth}
        \href{https://creativecommons.org/licenses/by-nc-nd/4.0/}{%
          \includegraphics[height=2\baselineskip]{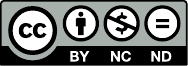}
        }
        \par 
        \kern 2pt
        \href{https://creativecommons.org/licenses/by-nc-nd/4.0/}{This work is licensed under a Creative Commons Attribution-NonCommercial-NoDerivatives 4.0 International License.}
      \end{minipage}
    }%
    \ifx\@tnotes\@empty\else\@tnotes\fi
  }
  {\typeout{*** Local no-indent CC license block successfully injected. ***}}
  {\errmessage{!!! Failed to patch \string\printFirstPageNotes for the license block!!!}}
\makeatother

\makeatletter
\xdef\myfootertxt{This is the author's accepted manuscript for an article accepted for publication in \@journal.}
\myfooter[L]{\myfootertxt}
\makeatother

\title{Advanced LPeg Techniques: A Dual Case Study Approach}

\author{Zixuan Zhu\,\orcidlink{0009-0009-4400-1682}\corref{cor1}}
\ead{10zhuzixuan@163.com}
\cortext[cor1]{Corresponding author at: College of Management Science and Information Technology, Hebei University of Economics and Business, China.}
\affiliation{organization={Hebei University of Economics and Business},
            country={China}}

\begin{abstract}
This paper presents advanced optimization techniques for Lua Parsing Expression Grammars (LPeg) through two complementary case studies: a high-performance JSON parser and a sophisticated Glob-to-LPeg pattern converter. We demonstrate how strategic grammar construction can dramatically improve parsing performance without modifying the underlying LPeg library. For the JSON parser, we implement substitution capture and table construction optimization to reduce memory allocation overhead and improve object processing. For the Glob converter, we introduce segment-boundary separation, implement Cox's flattened search strategy, and develop optimized braced condition handling to prevent exponential backtracking. Comprehensive benchmarks demonstrate that our JSON parser achieves processing speeds up to 125 MB/s on complex documents, consistently outperforming dkjson and showing competitive results against rxi\_json across most test cases. Our Glob-to-LPeg converter exhibits 14-92\% better performance than Bun.Glob and runs 3-14 times faster than Minimatch across diverse pattern matching scenarios. This research provides practical optimization techniques for LPeg-based parsers, contributing valuable strategies to the text processing ecosystem.
\end{abstract}


\end{frontmatter}

\section{Introduction}
\label{introduction}

Text parsing and pattern matching are critical in modern software development, underpinning tasks like processing configuration files, interpreting data formats, and crafting domain-specific languages. LPeg, a Lua-based implementation of Parsing Expression Grammars (PEG), combines PEG's expressive power and predictability with Lua's simplicity and adaptability in this domain. Academic research, such as Ierusalimschy (2009), delves into LPeg's internals as a PEG-driven pattern matching tool \citep{Ierusalimschy2009} while Medeiros (2008) dissects the parsing machine that drives its functionality \citep{medeiros2008parsing}. Despite its robust design, LPeg's advanced capabilities remain underutilized in practice.

This paper demonstrates advanced LPeg techniques through a dual case study approach, highlighting methods that extend beyond basic pattern matching. By focusing on optimized grammar design, we aim to demonstrate how these methods enhance performance while preserving code clarity and readability. Our exploration centers on two intricate parsing challenges:

\begin{enumerate}
\item \textbf{High-performance JSON parser}: We implement a JSON parser that leverages advanced LPeg optimizations to achieve performance comparable to hand-optimized Lua parsers. This case study showcases techniques including table construction optimization and substitution capture.
\item \textbf{Sophisticated Glob-to-LPeg converter}: We develop a converter that transforms Glob patterns into equivalent LPeg patterns, showcasing LPeg's flexibility in handling complex pattern matching scenarios. This study illustrates how clear separation of pattern matching rules and strategic optimization choices can lead to both performant and maintainable code.
\end{enumerate}

These studies illustrate how different LPeg grammar designs, applied to identical parsing goals, yield diverse performance outcomes when the library's strengths are fully leveraged. We reveal optimization techniques for PEG-based parsers that rely on thoughtful pattern construction rather than altering the core library. Our work offers actionable insights into building efficient parsers without sacrificing code quality.

The paper is organized as follows: We begin with an overview of LPeg's core concepts to build a basic understanding of LPeg; we also introduce existing JSON parsers and Glob matchers to provide context for our two case studies. Next, we dive into the case studies, detailing their implementation, optimization approaches, and technical hurdles. We then assess the performance and efficacy of our solutions, benchmarking them against existing alternatives. Finally, we distill our findings and explore how these LPeg techniques could apply to other parsing contexts.

Through these detailed, real-world examples, we aim to equip the Lua community with a deeper understanding of LPeg's capabilities, encouraging developers to harness its full power for complex text processing and parsing challenges. The full source is available at \url{https://github.com/brynne8/advanced-lpeg}.

\section{Related Work}
\label{related-work}

To lay the groundwork for exploring advanced LPeg applications, it is important to first review the technical background. This review encompasses not only the development of LPeg but also related parsing techniques, as well as studies on JSON parsers and Glob matchers.

\subsection{Overview of LPeg}
\label{overview-of-lpeg}

LPeg (Lua Parsing Expression Grammars) is a pattern matching library for the Lua programming language, offering powerful parsing capabilities. LPeg was first proposed by Roberto Ierusalimschy in 2008 as an efficient way to implement Parsing Expression Grammars (PEGs) in Lua. PEGs, introduced by Ford (2004) \citep{Ford2004}, provide an alternative to context-free grammars (CFGs) for specifying the syntax of programming languages and other structured data formats. One advantage of PEGs over CFGs is that a PEG specification not only describes the language syntax but also inherently defines how to parse it, eliminating the need for separate parsing algorithms or disambiguation rules that are often required with CFGs.

LPeg's design philosophy emphasizes simplicity, expressiveness, and efficiency. It allows developers to define grammars using a combination of pattern matching functions and operators, which can be easily composed to create complex parsing expressions. LPeg's core features include:

\begin{enumerate}
\item \textbf{Pattern Matching}: LPeg provides a concise and expressive syntax for defining patterns, enabling powerful and flexible pattern matching within Lua.

\item \textbf{Captures}: LPeg allows capturing and extracting specific parts of the input during parsing, enabling the construction of structured representations of the parsed data, such as abstract syntax trees (ASTs) or Lua tables. Among its various capture types:
   \begin{itemize}
   \item \textbf{Group captures} let you collect multiple related matches as a single unit
   \item \textbf{Match-time captures} provide dynamic control over the matching process by evaluating a function at runtime that can determine whether and how the match should proceed
   \item \textbf{Accumulator captures} modify previous captured values by applying a function that combines the last capture with the current one.
   \end{itemize}

\item \textbf{Grammar Definition}: LPeg grammars are defined using a set of rules, each associating a name with a parsing expression. These rules can be recursive, allowing the specification of complex language constructs. However, LPeg does not support left recursion and will detect and report such patterns as errors.

\item \textbf{Efficiency}: LPeg is designed with performance in mind, featuring a fast pattern matching engine implemented in C. It employs various optimization techniques, such as special purpose instructions, tail call optimization and head-fail optimization, to minimize parsing overhead.
\end{enumerate}

One of LPeg's key strengths lies in its seamless integration with Lua. LPeg grammars can be defined using Lua's expressive and dynamic features, such as first-class functions and table constructors. This enables developers to create modular and reusable parsing components that can be easily combined and extended.

LPeg has found widespread adoption in the Lua community for a variety of parsing tasks, including configuration file processing, data format parsing, and the implementation of domain-specific languages (DSLs). Its simple yet powerful API and efficient implementation have made it a go-to choice for many Lua developers.

In the following sections, we will explore advanced LPeg techniques and optimizations through two in-depth case studies: a high-performance JSON parser and a sophisticated Glob-to-LPeg converter. These case studies will demonstrate how LPeg can be leveraged to tackle complex parsing challenges while achieving high performance and strong expressiveness.

\subsection{JSON Parsers}
\label{json-parsers}

JSON parsing efficiency is crucial in modern software development, where applications must handle increasingly data-rich operations. The challenges in implementing an efficient JSON parser extend beyond simple text processing, encompassing the handling of deeply nested structures, dynamic type management, and memory optimization. These challenges have led to various implementation approaches across programming languages.

In various programming languages, there are dedicated JSON parsing libraries, such as \texttt{JSON.parse} in JavaScript, the \texttt{json} library in Python, Jackson and Gson in Java, and RapidJSON in C++. These parsers are usually implemented as handwritten recursive descent parsers or state machines to achieve efficient parsing.

\begin{samepage}
In the Lua ecosystem, common JSON parsing libraries include:
\begin{enumerate}
\item \textbf{dkjson}: A pure Lua implementation of a JSON parser that supports parsing using LPeg.
\item \textbf{cjson}: A high-performance Lua-C library specifically designed for fast JSON parsing and generation.
\item \textbf{rxi\_json}: A lightweight JSON library for Lua that achieves very fast parsing performance in pure Lua.\footnote{\url{https://github.com/rxi/json.lua}}
\end{enumerate}
\end{samepage}

These existing JSON parsing libraries in the Lua ecosystem provide a solid foundation for understanding different approaches to JSON parsing, from pure Lua implementations to high-performance C libraries.

\subsection{Glob Matchers}
\label{glob-matchers}

Glob patterns are primarily used for file path matching and are widely used in Unix/Linux systems \citep{glob7}. They provide a concise syntax for matching file paths through wildcards and pattern expressions. While conceptually simple, implementing efficient and correct glob matching presents interesting challenges, particularly regarding standardization and performance.

The implementations of glob matchers vary across programming environments, including the built-in glob command in Unix/Linux, Python's \textbf{glob} module, Node.js libraries like \textbf{minimatch}, and Go's \textbf{filepath.Glob} function. From analyzing these implementations, two predominant approaches emerge: converting glob patterns to regular expressions, and implementing dedicated parsers.

The Lua ecosystem has limited options for Glob pattern matching. An early attempt, lua-glob-pattern \citep{luaglobpattern}, tried to provide basic file globbing by converting glob patterns to Lua patterns, but its functionality was constrained by the inherent limitations of Lua patterns. Subsequently, lua-glob \citep{luaglob} emerged as a notable implementation built with LPegLabel (an extension of LPeg). NeoVim also includes its own LPeg-based glob implementation, designed specifically for its environment \citep{neovim_glob}.

The emergence of Microsoft's Language Server Protocol (LSP) 3.17 specification \citep{LSP317} has begun to address the historical lack of formal standardization in glob implementations, with projects like NeoVim adopting this specification as a reference. These existing implementations, with their different approaches and performance characteristics, provide a valuable insights for understanding the challenges and opportunities in glob pattern matching.

\section{Notation and Conventions}

To make the LPeg patterns in this paper more readable and understandable, we adopt the same syntax used in LPeg.re (except in long strings, forms like \texttt{\textbackslash 0} or \texttt{\textbackslash 31} are escaped representing specific characters), a Lua module for LPeg that allows patterns to be written in a string format closer to standard PEG syntax. For readers who are more familiar with LPeg's Lua API functions (such as \texttt{lpeg.P}, \texttt{lpeg.R}, \texttt{lpeg.S}, etc.), please refer to the LPeg.re official documentation\footnote{LPeg.re official documentation: \url{https://www.inf.puc-rio.br/~roberto/lpeg/re.html}} for a detailed explanation of the relationship between the two syntaxes.

\begin{samepage}
In this paper, we will use the following notation conventions:

\begin{itemize}
\item \textbf{Repetition}: \texttt{p\^{}n} ($n$ times), \texttt{p\^{}+n} ($\geq n$ times), \texttt{p\^{}-n} ($\leq n$ times)
\item \textbf{Capture}:
  \begin{itemize}
  \item Simple capture: \texttt{\{ p \}}
  \item Anonymous group capture: \texttt{\{: p :\}}
  \item Named group capture: \texttt{\{:name: p :\}} (where \texttt{name} is the name of the capture group)
  \item Table capture: \texttt{\{| p |\}}
  \item Constant/predefined capture: \texttt{p -> function/query/string}
  \item Match-time capture: \texttt{p => function}
  \item Accumulator capture: \texttt{e >{}> function}
  \item Substitution capture: \texttt{\{$\sim$ p $\sim$\}}
  \end{itemize}
\end{itemize}
\end{samepage}

These conventions aim to make it easier for readers to understand and follow the examples and case studies discussed throughout the paper.

\subsection{Symbols}

Throughout this paper, we use consistent symbols to enhance readability:

\begin{itemize}
\item $p$ denotes a pattern, which depending on context may refer to a PEG pattern or a Glob pattern
\item $s$ refers to the starting rule in a grammar, particularly when a grammar has only one rule
\item $\varphi(p)$ denotes the PEG transformation of a Glob pattern $p$, resulting in an equivalent PEG expression
\end{itemize}

Note that the transformed $\varphi(p)$ could be a grammar with several rules, in which case $\varphi(p)$ refers to the call to the starting rule of that grammar, rather than inlining the entire grammar. Additionally, $\varphi(p)$ can be used within larger PEG expressions, combining with other PEG syntaxes.

This notation allows us to precisely describe the transformation process from Glob patterns to their equivalent LPeg implementations, which is central to our Glob-to-LPeg converter case study.

\section{Case Studies}

To showcase the application of advanced LPeg techniques and demonstrate their impact on parsing performance and pattern matching expressiveness, we have selected two representative case studies for in-depth analysis. These case studies, focusing on a high-performance JSON parser and a sophisticated Glob-to-LPeg converter, allow us to explore a range of LPeg optimization strategies and extensions in practical scenarios.

As a widely used data exchange format, the implementation of a JSON parser can effectively demonstrate LPeg's advantages and potential for optimization when handling structured data. The Glob-to-LPeg conversion, on the other hand, demonstrates LPeg's flexibility and powerful features in handling complex pattern matching problems. Through these two case studies, we can see not only the practical application of LPeg but also explore how to use LPeg's features to optimize performance and improve code quality.

Next, let's analyze these two case studies in detail.

\subsection{Case Study: JSON Parser - A Concise Example}

This section describes the implementation of a JSON parser using LPeg that complies with the ECMA-404 JSON specification \citep{ecma404}. The discussion begins by presenting a slightly modified version of the Parsing Expression Grammar (PEG) description of JSON, as outlined by Yedidia \citep{Yedidia2021}.

\begin{listing}[ht]
\centering
\begin{minted}{text}
doc           <- JSON !.
JSON          <- __ (Number / Object / Array / String / True /
                     False / Null) __
Object        <- '{' (String ':' JSON (',' String ':' JSON)* /
                      __) '}'
Array         <- '[' (JSON (',' JSON)* / __) ']'
String        <- __ '"' StringBody '"' __
StringBody    <- ([^"\\0-\31]+ / Escape+)*
Escape        <- '\' (["{|\/bfnrt] / UnicodeEscape)
UnicodeEscape <- 'u' %x%x%x%x
Number        <- Minus? IntPart FractPart? ExpPart?
Minus         <- '-'
IntPart       <- '0' / [1-9][0-9]*
FractPart     <- '.' [0-9]+
ExpPart       <- [eE] [+-]? [0-9]+
True          <- 'true'
False         <- 'false'
Null          <- 'null'
__            <- %s*
\end{minted}
\caption{PEG grammar for JSON}
\label{lst:json-grammar}
\end{listing}

The grammar provided above is a concise representation of how one might typically define a JSON grammar for purposes of understanding and implementation. This string-based grammar can be directly utilized with LPeg's \texttt{re.compile} to produce a JSON validator. When JSON text conforms to this grammar, the validator returns the length of the JSON string plus one; for non-conforming text, it returns \texttt{nil}. This behavior is consistent with any grammar that does not include LPeg-specific captures.

To transform this validator into a full-fledged JSON parser, captures must be added to retain the JSON data as a Lua table structure. Consider the following examples for \texttt{Array} and \texttt{Number}:

\begin{minted}{peg}
Array  <- {| '[' ({: JSON :} __ (',' {: JSON :} __)* / __) ']' |}
Number <- ( Minus? IntPart FractPart? ExpPart? ) -> tonumber
\end{minted}

In this extended grammar, the \texttt{Array} rule captures JSON values within an outermost table capture, thereby aggregating them into a single Lua table. The \texttt{Number} rule uses the \texttt{tonumber} function to convert the matched string into a Lua number.

Furthermore, the \texttt{UnicodeEscape} rule cannot remain as currently defined. To correctly identify UTF-16 surrogate pairs, we need to handle two different cases:

\begin{minted}{peg}
UnicodeEscape <- 'u' -> '' (({[dD][89aAbB]%x%x} '\u' {%x%x%x%x})
                            -> surrogate / %x%x%x%x -> proc_uesc)
\end{minted}

Here, the \texttt{surrogate} function handles surrogate pairs, while \texttt{proc\_uesc} processes unicode characters in the Basic Multilingual Plane.

After incorporating these captures, the parser becomes capable of generating a usable Lua table structure from JSON data. However, the performance of this parser remains average, with potential risks of exceeding Lua's maximum stack limit.

\subsubsection{Optimizing Whitespace Handling}

Regarding the whitespace in this JSON grammar (the \texttt{\_\_} rule in the PEG), we have some ideas for improvement. One simple approach is to eliminate redundant matching of whitespaces. For the \texttt{JSON} rule, which handles whitespace uniformly at its end, we notice that the \texttt{String} rule also processes whitespace at its conclusion. We therefore consider removing the trailing whitespace part from the \texttt{String} rule, and adding the necessary whitespace after it when the \texttt{String} exists as an \texttt{Object} key. This transforms our \texttt{Object} rule to:

\begin{minted}{peg}
Object <- '{' (String __ ':' JSON (',' String __ ':' JSON)* / __) '}'
\end{minted}

Additionally, we realize that the trailing whitespace in the \texttt{JSON} rule affects the ability of its preceding choices to benefit from tail call optimization. Removing the trailing whitespace from the \texttt{JSON} rule will improve performance and reduce the LPeg VM stack usage for JSON parsing, helping to parse more deeply nested JSON files. The modified JSON rule becomes:

\begin{minted}{peg}
JSON <- __ (Number / Object / Array / String / True / False / Null)
\end{minted}

Corresponding locations that use the \texttt{JSON} rule would then supplement it with whitespace calls. Through these modifications, we ensure that the JSON grammar doesn't sacrifice performance in its formulation. This provides a solid foundation for discussing more advanced optimizations later.

\subsubsection{Accumulator vs Function-Based Object Construction}
LPeg's accumulator captures provide an intuitive approach to implementing JSON object parsing by enabling incremental construction of complex data structures. This mechanism effectively implements a fold operation within PEG, allowing us to aggregate captured values into a growing structure. Consider this implementation using accumulator captures:

\begin{minted}{peg}
Object <- '{' -> new_table ({: String __ ':' JSON :} >> add_prop __
          (',' {: String __ ':' JSON :} >> add_prop __)* / __) '}'
\end{minted}

The \texttt{add\_prop} function serves as the accumulator, incrementally constructing the object by adding each property. Here is its implementation, which clearly demonstrates how it processes the key-value pairs:

\begin{minted}{lua}
function add_prop(t, k, v)
  t[k] = v
  return t
end
\end{minted}

This implementation initializes an empty table via \texttt{new\_table}, which serves as the accumulator's initial value. As the parser encounters each key-value pair, it invokes \texttt{add\_prop} to incorporate the property into the growing table. The anonymous group capture ensures proper aggregation of the matched content, correctly binding both key and value parameters.

Considering LuaJIT's optimization characteristics, where well-crafted Lua code can potentially outperform C implementations, we can explore an alternative implementation strategy:

\begin{minted}{peg}
Object <- {| '{' (String __ ':' JSON __ (',' {: String __ ':' JSON :} __)*
           / __) '}' |} -> make_table
\end{minted}

This alternative approach employs a single function capture at the pattern's conclusion. By collecting all pairs first and constructing the object at once, it optimizes C-to-Lua transitions and simplifies group captures. The corresponding \texttt{make\_table} implementation requires careful attention:

\begin{minted}{lua}
function make_table(t)
  local len = #t
  local res = table.new(0, len / 2)
  for i=1, len, 2 do
    res[t[i]] = t[i + 1]
  end
  return res
end
\end{minted}

This implementation leverages strategic table management to enhance performance. Notably, it pre-allocates the result table with precise hash space dimensioning via \texttt{table.new} \citep{luajit_table_new}, enabling more efficient memory utilization by the Lua VM. This function-based approach demonstrates particular efficiency when processing JSON objects with numerous key-value pairs, where the benefits of bulk object construction become more pronounced.

\subsubsection{Substitution Capture}

Substitution capture, while less known, is a powerful feature of LPeg. PEG captures can be categorized into generative and substitutive. Generative captures build larger structures from captured components, while substitutive captures aim to replace parts of a string based on a predefined pattern.

LPeg's substitution capture falls into the latter category. Within a substitution capture, all captures replace the original matched content rather than generating results directly. This technique is particularly useful in scenarios like handling escapes within strings. Without substitution capture, the rules related to \texttt{String} might look like this:

\begin{minted}{text}
String        <- __ '"' {| StringBody |} -> fast_merge '"'
StringBody    <- ({ [^"\\0-\31]+ } / Escape+)*
Escape        <- '\' ({["{|\/]} / [bfnrt] -> str_esc / UnicodeEscape)
UnicodeEscape <- 'u' (({[dD][89aAbB]%x%x} '\u' {%x%x%x%x}) -> surrogate
                       / %x%x%x%x -> proc_uesc)
\end{minted}

Due to the presence of \texttt{Escape}, \texttt{StringBody} generates multiple string fragments. To merge these fragments efficiently, table capture is used to gather them into a Lua table, which is then merged using the \texttt{fast\_merge} function (actually \texttt{table.concat}\footnote{The \texttt{fast\_merge} alias is used to better emphasizing the intention of using this function.}). This approach is faster than concatenating strings individually, given Lua's inefficiency with the \texttt{..} operator for large-scale string concatenation. Using substitution capture simplifies and accelerates this process:

\begin{minted}{text}
String        <- __ '"' {~ StringBody ~} '"'
StringBody    <- ([^"\\0-\31]+ / Escape+)*
Escape        <- '\' -> '' (["{|\/] / [bfnrt] -> str_esc / UnicodeEscape)
UnicodeEscape <- 'u' -> '' (({[dD][89aAbB]%x%x} '\u' {%x%x%x%x})
                            -> surrogate / %x%x%x%x -> proc_uesc)
\end{minted}

Substitution capture optimizes the \texttt{getcaptures} phase by reducing memory allocation overhead, which is particularly beneficial for large JSON files. In generative capture, each fragment generates a Lua string, necessitating memory allocation and deallocation. Substitution capture, by contrast, uses \texttt{luaL\_Buffer} to merge C strings, significantly reducing the memory burden.

\subsection{Case Study: Glob-to-LPeg Converter - An In-Depth Analysis}
\label{case-study-glob-to-lpeg}

The JSON parser case study exemplifies a relatively straightforward application of LPeg optimization, providing a foundation for understanding basic concepts. However, the challenge of converting Glob patterns to LPeg representations introduces a more intricate problem, meriting a thorough exploration. This expanded case study enables us to delve into a wider array of LPeg features and optimization strategies within a practical context.

Glob patterns offer a concise yet powerful syntax for file path matching. The \texttt{vim.glob.to\_lpeg} converter in the NeoVim project \citep{neovim_glob} demonstrates this meta-level approach—using an LPeg grammar to parse Glob pattern syntax and, through its captures, produce equivalent LPeg patterns that can then match file paths. While this implementation highlights the potential of such conversions, it leaves room for improvement. In this section, we propose Peglob, a more comprehensive converter that employs the same fundamental approach but with enhanced functionality and precision.

\subsubsection{Overview of Glob Grammar}
\label{glob-spec}

We implement Glob patterns following the LSP 3.17 specification \citep{LSP317}, which defines patterns with various special characters and syntaxes, including:

\begin{itemize}
\item \texttt{*} to match zero or more characters in a path segment
\item \texttt{?} to match on one character in a path segment
\item \texttt{**} to match any number of path segments, including none
\item \texttt{\{\}} to group conditions (e.g. \texttt{*.{ts,js}} matches all TypeScript and JavaScript files)
\item \texttt{[]} to declare a range of characters to match in a path segment
  (e.g., \texttt{example.[0-9]} to match on \texttt{example.0}, \texttt{example.1}, \ldots)
\item \texttt{[!...]} to negate a range of characters to match in a path segment
  (e.g., \texttt{example.[!0-9]} to match on \texttt{example.a}, \texttt{example.b},
  but not \texttt{example.0})
\end{itemize}

The LSP specification provides a broad definition of Glob patterns without enforcing a strict grammar. To implement a well-defined PEG version of Glob patterns, we apply the following constraints:

\begin{itemize}
\item A Glob pattern must match an entire path, with partial matches considered failures.
\item The pattern only determines success or failure, without specifying which parts correspond to which characters.
\item A \textbf{path segment} is the portion of a path between two adjacent path separators (\texttt{/}), or between the start/end of the path and the nearest separator.
\item The \texttt{**} (\textbf{globstar}) pattern matches zero or more path segments, including intervening separators (\texttt{/}). Within pattern strings, \texttt{**} must be delimited by path separators (\texttt{/}) or pattern boundaries and cannot be adjacent to any characters other than \texttt{/}. If \texttt{**} is not the final element, it must be followed by \texttt{/}.
\item \texttt{\{\}} (\textbf{braced conditions}) contains valid Glob patterns as branches, separated by commas. Commas are exclusively used for separating branches and cannot appear within a branch for any other purpose. Nested \texttt{\{\}} structures are allowed, but \texttt{\{\}} must contain at least two branches—zero or one branch is not permitted.
\item In \texttt{[]} or \texttt{[!...]}, a \textbf{character range} consists of character intervals (e.g., \texttt{a-z}) or individual characters (e.g., \texttt{w}). A range including \texttt{/} won't match that character.
\end{itemize}

\subsubsection{Pattern Analysis and Implementation Strategy}

The key to efficient Glob pattern matching lies in a fundamental architectural decision: strictly separating in-segment pattern matching from segment-crossing behaviors. This separation simplifies both implementation and optimization while maintaining pattern matching correctness.

Within a path segment, patterns operate in a well-defined, constrained environment. Patterns in a segment can be categorized into two types: 

\begin{enumerate}
\item \textbf{Fixed-length patterns}: Including string literals, single character wildcards (\texttt{?}), and character classes (\texttt{[]})
\item \textbf{Variable-length patterns}: Including star (\texttt{*}) and globstar (\texttt{**})
\end{enumerate}

This classification guides our implementation strategy, particularly in handling path segments efficiently. Within path segments, patterns alternate between fixed-length \texttt{Word} constructs and variable-length \texttt{Star} operators, forming the basis of our matching algorithm.

This controlled scope of segment-level matching allows us to safely apply sophisticated optimizations without risking unintended interactions with segment-crossing patterns. The separation of concerns ensures that our optimization strategies remain effective while maintaining pattern matching correctness.

The handling of segment-crossing patterns, particularly the globstar (\texttt{**}) operator, becomes cleaner through this separation. The globstar operator is treated as a distinct element that operates purely at the segment boundary level. This clarity of purpose makes the implementation both more robust and more maintainable.

\paragraph{Segment-Level Pattern Matching}
\label{segment-level}

\begin{figure}[h]
\centering
\includegraphics[width=\textwidth]{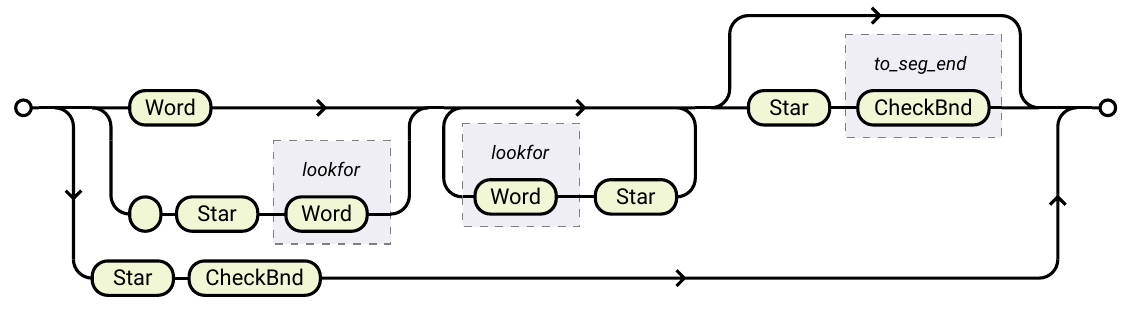}
\caption{Syntax of Segment. The gray boxes represent accumulator captures, and the nodes inside them show the applied pattern.}
\label{fig:segment-syntax}
\end{figure}

In path segment matching, patterns alternate between \texttt{Word} and \texttt{Star}, as outlined in the syntax diagram at Figure~\ref{fig:segment-syntax}. Both \texttt{lookfor} and \texttt{to\_seg\_end} operate as accumulator captures, folding over the matched Glob pattern to incrementally build the result. The \texttt{Segment} grammar splits into two branches: the first begins with \texttt{Word}, and the second starts with an empty string (denoted by an empty circle in the railroad diagram); both serve as the initial accumulator value.

A constraint in this syntax is \texttt{CheckBnd}, enforced whenever a \texttt{Star} appears at the end of a \texttt{Segment} match:

\begin{minted}{peg}
CheckBnd <- &'/' / !.
\end{minted}

This rule prevents globstar patterns from erroneously spanning segment boundaries by requiring either a forward slash (\texttt{/}) or the end of input. By doing so, it confines star matching to within a single segment, ensuring precise control over pattern scope.

The foundation of fixed-length pattern matching lies in the \texttt{Word} rule:

\begin{minted}{peg}
Word    <- !'*' {| FIRST WordAux |}
WordAux <- Branch / (Simple+ Branch?)
Simple  <- Token+ Boundary?
Token   <- Ques / Class / Escape / Literal
\end{minted}

Here, \texttt{Word} establishes the basis for matching fixed-length patterns. An optimization comes from the \texttt{FIRST} rule, which extracts a deterministic first character when possible:

\begin{minted}{peg}
FIRST <- {:F: '' => get_first :}
\end{minted}

\begin{minted}{lua}
function get_first(s, i)
  if letter:match(s, i) then return true, s:sub(i, i)  
  else return false end
end
\end{minted}

Due to LPeg's limitations, extracting the first character requires match-time capture, as other capture types cannot capture the character without consuming it.

Captured via LPeg's named group mechanism, this first-character information drives the \texttt{lookfor} function, which generates tailored LPeg search patterns. When a deterministic first character is identified, \texttt{lookfor} produces an optimized pattern ($s_\text{opt}$); otherwise, it defaults to a plain pattern ($s_\text{plain}$):

\begin{align*}
s_\text{plain} &\leftarrow p \; / \; . \; s_\text{plain} \\
s_\text{opt}   &\leftarrow p \; / \; . \; (!\text{first}(p) \; .)^* \; s_\text{opt}
\end{align*}

The \texttt{lookfor} function's ability to produce $s_\text{opt}$ is key to its efficiency. When a deterministic first character is identified, $s_\text{opt}$ optimizes the search by skipping positions in the input where the current character doesn't match the pattern $p$'s required starting character. For example, when searching for pattern ``cat'', the optimized rule becomes 

\begin{center}
\verb|s <- "cat" / . [^c]* s|
\end{center}

where \texttt{[\textasciicircum c]*} efficiently skips all non-`c' characters after each failed match attempt. This \textbf{skipping false starts} technique, adapted from Ierusalimschy's work \citep{Ierusalimschy2009}, boosts performance by avoiding unnecessary matching attempts, unlike the slower $s_\text{plain}$, which lacks this optimization.

Handling star patterns within segments poses a distinct challenge. Drawing on Cox's insight into pattern matching structure \citep{Cox2017}, we avoid nested matching—which often leads to exponential backtracking—and instead flatten the \texttt{lookfor} function calls. For instance, a pattern like \texttt{a*b?c*x} is processed as sequential calls—\texttt{a lookfor(b?c) lookfor(x)}—rather than a nested structure like \texttt{a lookfor(b?c lookfor(x))}. This flattening preserves correctness while significantly reducing complexity. The \texttt{lookfor} function's accumulator capture maintains state across the matching process, and its use of $s_\text{opt}$ ensures that the search for each subsequent pattern benefits from the false-start-skipping optimization tailored to Glob semantics.

For star patterns at segment ends, the \texttt{to\_seg\_end} function translates them into a greedy match (\texttt{[\textasciicircum/]*}), appended to the existing grammar. Since the star is terminal, it consumes all remaining characters within the segment boundary, with \texttt{CheckBnd} ensuring no overreach beyond the slash or input end.

The \texttt{Word} rule also integrates the \texttt{Branch} rule to support braced conditions, introducing variable-length elements. While this departs from strict fixed-length matching, it raises complexities addressed later in Section \ref{braced-conditions}.

Together, these components—the deterministic character extraction of \texttt{FIRST}, the pattern generation and optimization of \texttt{lookfor}, and Cox's flattened search strategy—form a robust framework for segment-level pattern matching. By enforcing strict segment boundaries and leveraging these optimizations, the system efficiently handles complex patterns while maintaining semantic accuracy.

\paragraph{Cross-Segment Pattern Matching}

With segment-level matching established as a foundation, we can construct higher-level grammatical structures. In our Glob grammar, the topmost layer consists of the \texttt{Glob} and \texttt{Element} rules, as shown in Listing~\ref{lst:glob-grammar}. The \texttt{Glob} rule handles both absolute paths (beginning with slash) and local paths (beginning with \texttt{Element}). Consequently, the \texttt{Glob} rule starts with an optional \texttt{Element}.

\begin{listing}[h]
\begin{minted}{peg}
Glob    <- Element? (slash Element)* (slash? eof)
Element <- DSeg / DSEnd / Segment (slash Segment)*
                                  (slash eof / eof?)
DSeg    <- '**/' (Element / eof)
DSEnd   <- '**' !.
\end{minted}
\caption{The topmost rules of our Glob-to-LPeg converter, Peglob. The captures are removed to present a clear grammar. \texttt{DSeg} and \texttt{DSEnd} rules are used to process globstars.}
\label{lst:glob-grammar}
\end{listing}

The rules \texttt{Glob}, \texttt{Element}, and \texttt{DSeg} each incorporate EOF handling to ensure proper termination: EOF functions as a boundary marker for these rules. The first branch of the \texttt{Element} rule ending (\texttt{slash eof / eof?}) is deliberately designed this way to ensure that a slash not followed by EOF is handled by the \texttt{(slash Element)*} portion of the \texttt{Glob} rule. Each rule maintains strict boundary awareness, avoiding consumption of input required by other rules.

\texttt{Element}, functioning as a fundamental matching unit, provides an abstraction that can be repeated at the higher level (\texttt{Glob} rule) while integrating globstar patterns with ordinary \texttt{Segment} groups (structured as \texttt{Segment (slash Segment)*}). Regarding the downward functionality, we must first examine how globstar semantics are implemented in our PEG grammar.

We categorize globstars into two types based on their position: globstars appearing at the pattern's end are processed by the \texttt{DSEnd} (double-star ending) rule, as they do not involve searching for subsequent patterns. For globstars in other positions, we employ the \texttt{DSeg} (double-star segment) rule, which processes the globstar together with the following slash, forming a segment-like structure (though its function is inherently cross-segment).

This differentiation enables simpler handling of straightforward cases. The \texttt{DSEnd} rule directly translates a terminal globstar into a greedy match (\texttt{.*}), which consumes all remaining input. The \texttt{DSeg} processing is more complex, transforming the globstar and its continuation (the \texttt{Element} or EOF matched in the rule) into a search structure. For a pattern like \texttt{**/p}, the resulting PEG takes the form:

\[
s \leftarrow \varphi(p) \; / \; \verb|[^/]*| \; \verb|'/'| \; s
\]

While this search represents a form of lazy matching for $\varphi(p)$, its second branch efficiently skips in-segment content since we only need to determine whether globstar's continuation can be matched from a certain segment head. By restricting the globstar's continuation to a single \texttt{Element}, we mirror the flattened search strategy discussed earlier. By enabling \texttt{Element} to match \texttt{Segment} groups, \texttt{DSeg}'s search need only concern itself with matching this portion, without considering subsequent glob pattern matching.

The use of \texttt{Segment} groups rather than individual \texttt{Segment}s addresses an important edge case. Consider a Glob pattern like \texttt{**/b/d/**} and a path \texttt{/a/b/c/b/d/e}. If the last \texttt{Element} branch matched only a single \texttt{Segment}, after matching \texttt{/a/b}, the grammar would expect \texttt{/d}, but encountering \texttt{/c} would cause the match to fail since the prior \texttt{/b} cannot be backtracked.

This design allows our glob matcher to avoid backtracking at the cross-segment level. Combined with segment-level optimizations, this establishes a robust foundation for efficient pattern matching across both segment and path levels.

\subsubsection{Braced Conditions and Expansion}
\label{braced-conditions}

Braced conditions (\texttt{\{\}}) introduce branching complexity that requires careful handling. While our initial approach aimed to handle complex structures without expansion, practical considerations led us to implement brace expansion as a preprocessing step.

Brace expansion, a Unix shell feature, simplifies input by specifying similar string parameters. For example, \texttt{enable\_\{audio,video\}} expands to \texttt{enable\_audio} and \texttt{enable\_video}. The expansion rules are as follows: comma-separated segments within braces represent alternatives, generating all possible combinations. These alternatives accumulate in a Cartesian product and expand in left-to-right order. Unmatched braces and special characters are treated as literals \citep{rosetta}.

Before delving deeper, let's examine how braced conditions are transformed into PEG. Braced conditions in Glob patterns are analogous to alternation in regular expressions. Therefore, our approach to handling braced conditions mirrors Medeiros's \citep{Medeiros2011} treatment of alternation. For Glob patterns $R$, $S$, and $T$, the expression $\{R,S\}T$ transforms into PEG form following this derivation:

\begin{align*}
  \varphi(\{R,S\}T) \Rightarrow \varphi(R T) \; / \; \varphi(S T)
\end{align*}

For instance, the glob pattern \texttt{\{foo,bar\}baz} transforms into the PEG \verb|'foo' 'baz' / 'bar' 'baz'|. To further illustrate this concept, consider a more complex example: the glob pattern \texttt{\{ab,c\{d,e\}\}fg}. We first expand the braced conditions, then combine them with the subsequent pattern:

\[
\verb|{ab,c{d,e}}fg| \Rightarrow
\begin{pmatrix}
\verb|ab| \\
\verb|cd| \\
\verb|ce|
\end{pmatrix} 
\verb|fg| \Rightarrow
\verb|'ab'|\;\verb|'fg'| \; / \; \verb|'cd'| \;\verb|'fg'| \; / \;\verb|'ce'| \; \verb|'fg'|
\]

This means that when we transform braced conditions into PEG, we must consume the full pattern that follows the braced condition, to construct the equivalent ordered choices structure in PEG.\footnote{Even with Section \ref{opt-group-condition} optimizations, we consume at least to the segment's end} We've designed a two-level PEG structure to handle braced condition transformations.

At the lower level, the \texttt{CondList} and \texttt{Cond} grammar rules work together to implement true Unix shell-style brace expansion:

\begin{minted}{peg}
CondList <- ('{' Cond (',' Cond)+ '}')
Cond     <- ([^,{}]+ / CondList)+ / ε
\end{minted}

Here, $\varepsilon$ denotes an empty string. By folding captures on \texttt{CondList} and \texttt{Cond}, we derive an expanded list of strings in a table format.

At a higher level, these expansion results are concatenated and transformed into LPeg through \texttt{concat\_tail} in the \texttt{Branch} rule:

\begin{minted}{peg}
Branch <- (CondList {.*}) -> concat_tail
\end{minted}

This covers the basic processing logic for braced conditions. In the glob grammar, braced conditions are embedded within the \texttt{Word} rule through the \texttt{WordAux} rule. This design choice is critical because braced conditions can vary in scope—either staying within a single path segment or extending across multiple segments—depending on their internal patterns. By integrating them into \texttt{WordAux}, the grammar accommodates these differences seamlessly, without requiring significant changes to its core structure.

Moreover, this integration ensures that braced conditions are handled consistently with standard \texttt{Word} patterns, even in segments that include the star (\texttt{*}). This consistency allows the \texttt{lookfor} function to operate reliably across all cases. As a result, the grammar correctly parses braced conditions without complicating its structure.

\paragraph{Corner Cases with Star and Globstar}

The interaction between brace expansion and star/globstar patterns creates several important corner cases:

\begin{itemize}
\item \textbf{Scenario 1}: \texttt{...*\{*/p,...\}}, where the star before the brace and stars within brace branches could potentially merge into a globstar.
\item \textbf{Scenario 2}: \texttt{...**\{*p,...\}}, where the globstar takes precedence over the star, converting into three tokens: \texttt{**}, \texttt{*}, and \texttt{p}.
\item \textbf{Scenario 3}: \texttt{...**\{**/p,...\}}, which, regardless of expansion, results in two globstars followed by \texttt{/p}.
\item \textbf{Scenario 4}: \texttt{...q\{/**/p,...\}}, where \texttt{q} is a \texttt{Word} not ending in a star or globstar. This scenario expands into a valid pattern (\texttt{q/**/p}).
\end{itemize}

Our handling of these cases ensures consistent behavior while maintaining the system's overall integrity:

\begin{enumerate}
\item \textbf{For Scenario 1}: star merging is prevented, because we will treat the pattern starting from \texttt{\{} as a whole by \texttt{Branch} rule, embedded into \texttt{Word}, as explained before. While this approach may differ from some other Glob matchers, such patterns are rare in practice.
\item \textbf{For Scenario 2 and 3}: these cases are avoided by our grammar constraints, as they would produce invalid patterns according to our globstar positioning rules.
\item \textbf{For Scenario 4}: valid combinations with path separators match correctly, maintaining consistent matching semantics even with variable-length content.
\end{enumerate}

\paragraph{Performance Optimization for Braced Conditions}
\label{opt-group-condition}

Embedding braced conditions within \texttt{Word} can create expansive \texttt{Word} patterns with variable-length content, clashing with the original intent of \texttt{Word} rules meant for fixed-length patterns. When the \texttt{lookfor} function searches for such a pattern, it might have to scan to the match's end, since each branch in the PEG transformation includes content from the current position to EOF. This inefficiency pushes for an optimization to trim down the tail content in each branch, especially when certain conditions make the trailing part unnecessarily heavy, while keeping the transformation equivalent.

To address this, we propose an optimization strategy for specific scenarios where:

\begin{itemize}
\item \textbf{Constraint 1}: After the pattern is brace expanded, no branch within the braced condition contains \texttt{/} or \texttt{**}.
\item \textbf{Constraint 2}: The ``tail'' (all characters after the braced condition) has a prefix ending with \texttt{/}. This prefix, excluding its last character, contains neither \texttt{/}, \texttt{\{}, nor \texttt{**}.
\end{itemize}

Under these constraints, we can transform the pattern matching to reduce the search space while maintaining correctness:

\begin{align*}
&A [\varphi(B_1 T) \; / \; \varphi(B_2 T) \; / \; ...  \; / \; \varphi(B_n T)] \\
\Rightarrow &A [\varphi'(B_1 P) \; / \; \varphi'(B_2 P) \; / \; ...  \; / \; \varphi'(B_n P)] \varphi(Q)
\end{align*}

Where:
\begin{itemize}
\item $\varphi(p)$ represents the corresponding PEG of the Glob pattern $p$
\item $\varphi'(p)$ converts to PEG and transforms EOF matches into \verb|&'/'| lookahead predicates
\item $A$ is the fixed-length pattern before the grouping condition
\item $B_1, B_2, ..., B_n$ are the expanded branches  
\item $T$ is the tail string
\item $P$ is the prefix of the tail string meeting the constraints, with the last \texttt{/} character removed
\item $Q$ is the remainder of $T$ after removing $P$
\end{itemize}

The transformation maintains correctness because:
\begin{enumerate}
\item No branches contain \texttt{/} or \texttt{**}, so grouping conditions don't cross segment boundaries
\item Constraint 2 ensures $P$ doesn't cross segment boundaries
\item $B_i P$ remains within a segment
\item The successful match of $Q$ is independent of branch selection
\item At least one $B_i P$ matches the character before the first \texttt{/} 
\end{enumerate}

\begin{figure}
\centering
\includesvg[width=0.6\textwidth]{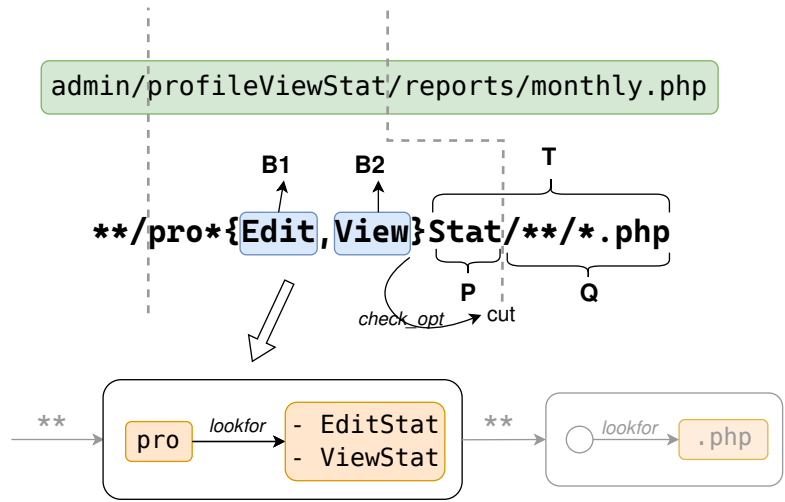}
\caption{Glob pattern with the path to match on the top, and the transformed syntax (diagram depiction) below. $B_1$, $B_2$ and $T$, $P$, $Q$ correspond to the patterns marked in the diagram. Empty circle means matching empty string in the diagram.}
\label{fig:opt-group-diagram}
\end{figure}

To implement this optimization in braced conditions, we need to modify the previously defined \texttt{Branch} rule. Since we need to dynamically determine the matching length of the \texttt{Branch} rule, we utilize LPeg's match-time capture. The modified \texttt{Branch} rule becomes:

\begin{minted}{peg}
Branch <- { CondList } => check_opt -> concat_tail
\end{minted}

The added \texttt{check\_opt} function evaluates the braced condition and subsequent tail against the two constraints mentioned above. If the conditions are met, it identifies the cut position to split the tail string $T$ into $P$ and $Q$. This match-time capture will only consume up to the cut position rather than continuing to EOF, allowing the \texttt{concat\_tail} function to merge only the $B_i P$ portion.

Figure~\ref{fig:opt-group-diagram} illustrates this concept with a concrete example. The middle section shows a Glob pattern containing a braced condition. After matching the braced condition \texttt{\{Edit,View\}}, \texttt{check\_opt} immediately verifies whether the branches \texttt{Edit} and \texttt{View} satisfy Constraint 1, and whether the prefix of $T$ (in this case, \texttt{Stat/}) meets Constraint 2. Upon confirmation, it determines the cut position, dividing $T$ into $P$ and $Q$, before invoking \texttt{concat\_tail} to generate the ordered choice grammar structure.

The key insight is that by separating at this cut position, we ensure that the brace expansion only needs to process up to the next segment boundary, rather than the entire remaining pattern. This strategic truncation prevents exponential growth of branch content while maintaining semantic correctness, transforming what could be an expensive operation into a manageable one.

\section{Evaluation}
\label{evaluation}

To comprehensively evaluate the effectiveness of the LPeg optimization techniques and the Glob-to-LPeg conversion method proposed in this paper, we designed a series of experiments. These experiments aim to test the performance of our JSON parser and Glob-to-LPeg converter from different angles and compare them with existing solutions. All experiments were conducted in a controlled environment using an Intel Core i7-8550U processor, 8GB DDR4 2400MHz RAM, running Alpine Linux v3.18 with LuaJIT 2.1.0-beta3 and LPeg 1.1.0. Each benchmark included 3 warmup runs to account for JIT behavior and was executed 10 times to ensure statistical significance.

\subsection{JSON Parser Performance}
\label{json-parser-performance}

\begin{figure}
\centering
\includesvg[width=\textwidth]{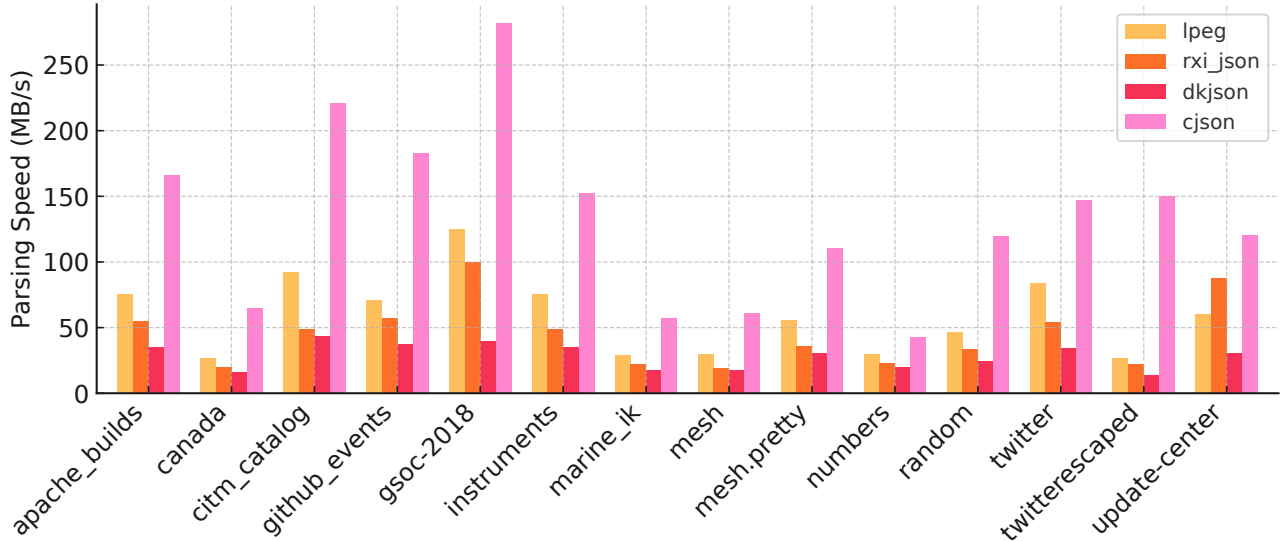}
\caption{Parsing speeds (in MB/s) of various JSON parsers}
\label{fig:json-parser-benchmark}
\end{figure}

We evaluated our optimized LPeg JSON parser's performance against existing JSON parsers (cjson, dkjson, and rxi\_json) using a diverse set of JSON files ranging from 64KB to 3MB with varying structural complexities from \cite{langdale2019parsing}. As shown in Figure~\ref{fig:json-parser-benchmark}, our implementation demonstrated competitive performance across a spectrum of document characteristics.

Our parser demonstrated efficiency across various data compositions. For instance, with string-heavy documents like gsoc-2018.json (34,128 strings), it achieved 125.00 MB/s. For number-intensive files such as marine\_ik.json (over 245,000 numerical entries) and numbers.json (almost exclusively float values), speeds were 29.31 MB/s and 29.76 MB/s respectively. This efficiency was also evident in mixed-content files. Notably, on citm\_catalog.json (rich in objects and strings) and instruments.json (mixed strings, integers, objects), our parser achieved 92.12 MB/s and 75.32 MB/s respectively. These speeds significantly surpassed rxi\_json's performance, which processed citm\_catalog.json at 48.83 MB/s and instruments.json at 48.64 MB/s. Our parser also effectively processed other mixed-type documents like github\_events.json at 70.58 MB/s.

While the C-based cjson consistently maintained its performance lead due to its lower-level implementation, our LPeg-based parser consistently outperformed dkjson in all benchmarks. Furthermore, it demonstrated strong competitiveness against rxi\_json, surpassing its speed in 13 out of the 14 test cases. The only exception was update-center.json, where rxi\_json (87.52 MB/s) was faster than our parser (59.82 MB/s), even though update-center.json has a high string content (27,229 strings); its particular structural characteristics appeared to favor rxi\_json's approach. These results validate our optimization strategy and demonstrate the viability of LPeg-based parsing for performance-critical JSON processing tasks.

\begin{figure}[h]
\centering
\includesvg[width=\textwidth]{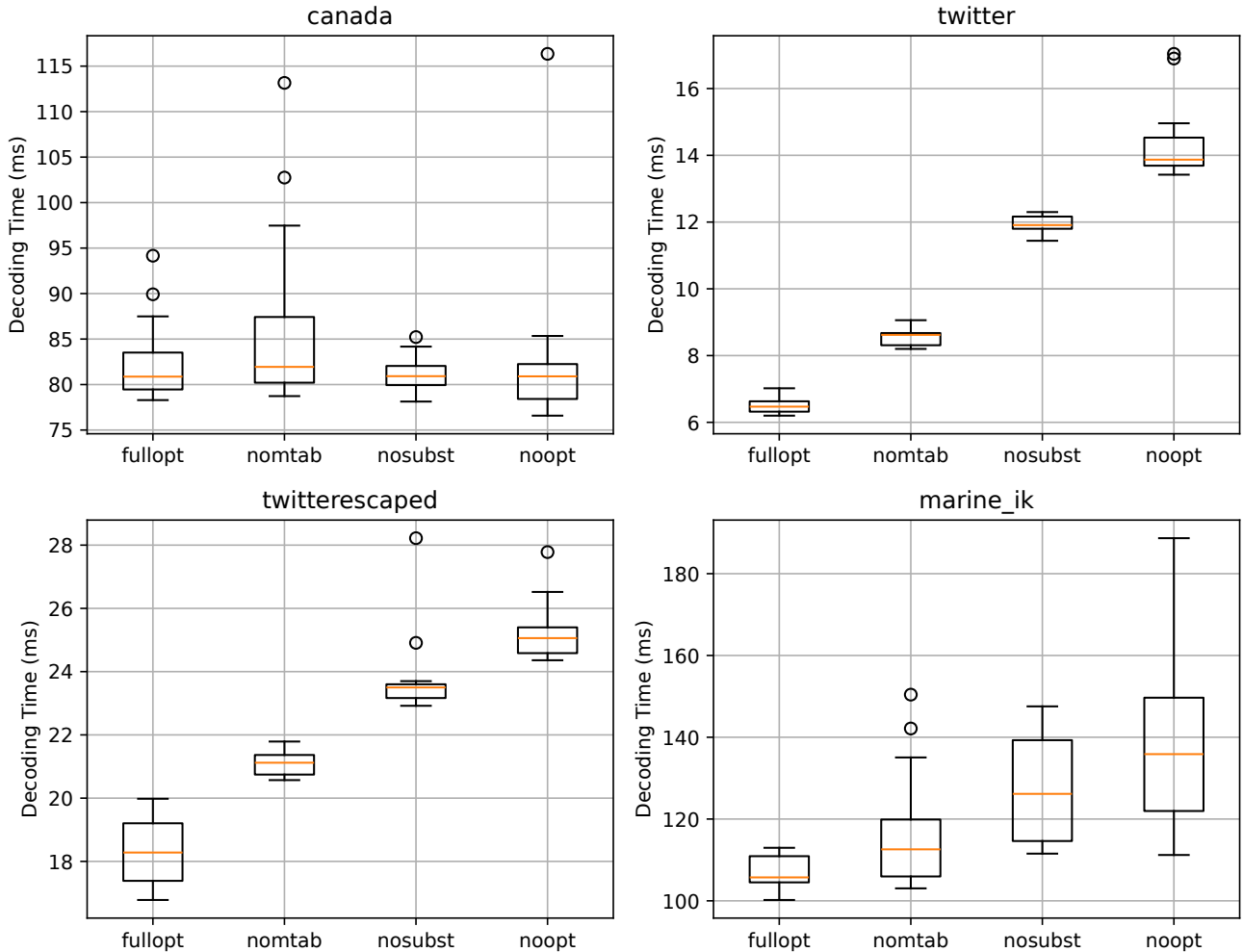}
\caption{Impact of optimization techniques on JSON parsing performance. Configurations: \texttt{fullopt} (fully optimized), \texttt{nomtab} (without table construction optimization), \texttt{nosubst} (without substitution capture), \texttt{noopt} (no optimizations).}
\label{fig:optimization-impact}
\end{figure}

We also analyzed the contribution of different optimization techniques to overall performance by evaluating the impact of two key optimizations: table creation (via \texttt{make\_table}) and substitution capture. Four diverse datasets were considered—canada.json for numerical processing, twitter.json for Unicode handling, twitterescaped.json for escape sequence processing, and marine\_ik.json for complex nested structures—with performance measured as the median decoding time under four configurations: fully optimized, without table creation, without substitution capture, and with no optimizations, as depicted in Figure~\ref{fig:optimization-impact}.

For the string-heavy datasets, the benefits of these optimizations were pronounced. In twitter.json, the fully optimized median was 6.43 ms; however, disabling table construction optimization increased this to 8.62 ms (about a 34\% slowdown), while omitting substitution capture increased the time to 11.91 ms (an increase of roughly 85\%). Similarly, for twitterescaped.json, the fully optimized median of 18.89 ms rose to 21.12 ms without table construction optimization (an 11.8\% increase) and to 23.50 ms without substitution capture (around a 24.4\% slowdown).

In contrast, canada.json exhibited minimal sensitivity to these optimizations—with median times varying by less than 2.5\%—indicating limited benefit for numerical data. The marine\_ik.json dataset, while structurally complex, showed moderate performance degradation: its fully optimized median of 105.73 ms increased to 112.59 ms (roughly a 6.5\% slowdown) when table creation was disabled and to 126.16 ms (about a 19.3\% increase) without substitution capture, reaching 135.87 ms when both were removed. These findings underscore that optimization effectiveness is highly data-dependent, with string-heavy JSON benefiting significantly from substitution capture, numerically-dominated content showing negligible response to optimizations, and structurally complex data exhibiting moderate but meaningful improvements.

\subsection{Glob-to-LPeg Converter Evaluation}
\label{glob-to-lpeg-evaluation}

We evaluated the correctness and performance of our Glob-to-LPeg converter, Peglob, focusing on how our architectural decisions affected real-world usage. Using the Bun.js Glob test suite as our foundation, we categorized test cases into six groups: basic patterns, star patterns, globstar patterns, brace expansions, Unicode patterns, and extension patterns. We selected this test suite for its comprehensive coverage of edge cases and complex pattern interactions, including intricate combinations of stars, globstars, character classes, and brace expansions.

We made two specific modifications to the test suite:
\begin{enumerate}
\item Removed tests with invalid surrogate pairs that would trigger exceptions in Lua's string handling
\item Excluded bash-specific extension tests that fall outside our formal Glob grammar specification
\end{enumerate}

Correctness was assessed using \textbf{accuracy} (proportion of correct matches) and \textbf{precision} (proportion of true matches among claimed true matches) for each test case category. These are calculated as:

\begin{itemize}
\item \textbf{Accuracy}: $(\text{TP} + \text{TN}) / (\text{TP} + \text{TN} + \text{FP} + \text{FN})$
\item \textbf{Precision}: $\text{TP} / (\text{TP} + \text{FP})$
\end{itemize}

Here, \textbf{TP} (true positives), \textbf{TN} (true negatives), \textbf{FP} (false positives), and \textbf{FN} (false negatives) are standard evaluation metrics.

\begin{figure}[h]
\centering
\includesvg[width=\textwidth]{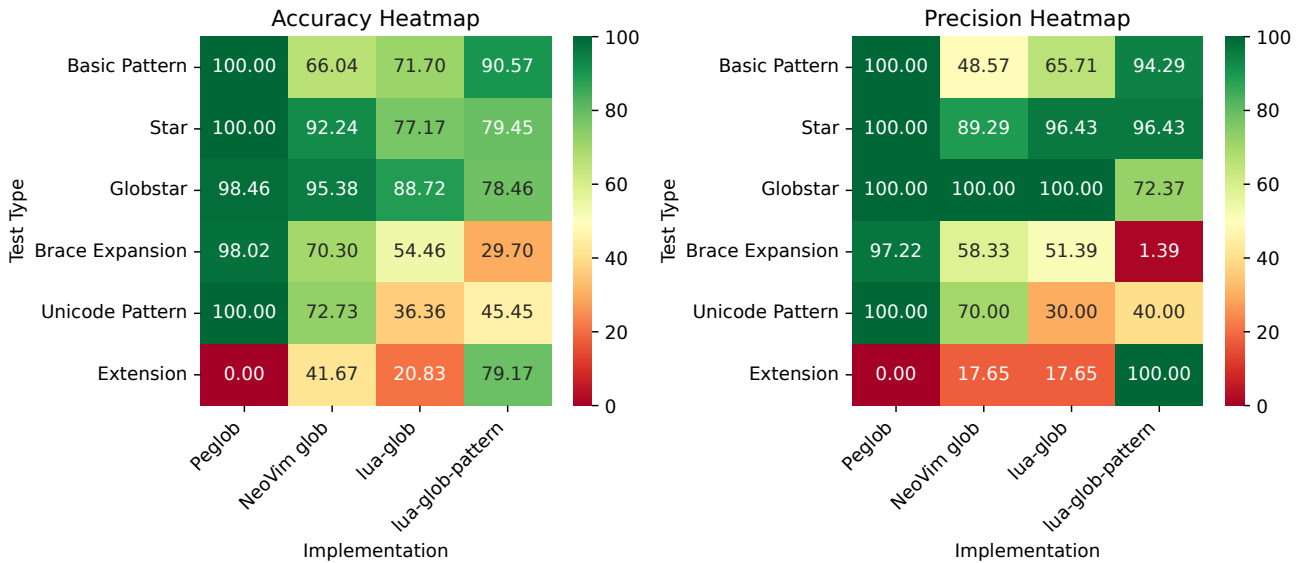}
\caption{Correctness of glob matchers}
\label{fig:glob-convert-example}
\end{figure}

As shown in Figure~\ref{fig:glob-convert-example}, Peglob demonstrates exceptional reliability across standard glob patterns, achieving perfect 100\% accuracy and precision in basic patterns, star patterns, and Unicode handling. Its robustness extends to complex cases with impressive results in globstar patterns (98.46\% accuracy, 100\% precision) and brace expansions (98.02\% accuracy, 97.22\% precision). These results validate our architectural decisions, particularly our strict grammar specification which deliberately rejects all extension patterns—ensuring predictable behavior for standard glob patterns at the expense of specialized extensions that allow \texttt{**} within path segments.

Comparative analysis reveals distinct trade-offs among existing implementations. Lua-glob-pattern excels with extension patterns (100\% precision, 79.17\% accuracy) but significantly underperforms with brace expansions (1.39\% precision), revealing fundamental limitations in Lua's pattern matching. The LPeg-based implementations show varying specializations: NeoVim's matcher handles globstar patterns well (95.38\% accuracy, 100\% precision) but struggles with basic patterns (66.04\% accuracy), while lua-glob achieves perfect precision in globstar matching (100\%) but falls short in overall consistency, particularly with Unicode (36.36\% accuracy). This performance landscape highlights the challenge of balancing comprehensive glob feature support with reliable pattern interpretation—a balance where Peglob's grammar-driven approach demonstrates clear advantages for standard use cases.

\begin{listing}
\centering
\begin{minted}[linenos,breaklines,breakanywhere]{text}
{src,extensions}/**/test/**/{fixtures,browser,common}/**/*.{ts,js}
{extensions,src}/**/{media,images,icons}/**/*.{svg,png,gif,jpg}
{.github,build,test}/**/{workflows,azure-pipelines,integration,smoke}/**/*.{yml,yaml,json}
src/vs/{base,editor,platform,workbench}/test/{browser,common,node}/**/[a-z]*[tT]est.ts
src/vs/workbench/{contrib,services}/**/*{Editor,Workspace,Terminal}*.ts
{extensions,src}/**/{markdown,json,javascript,typescript}/**/*.{ts,json}
**/{electron-sandbox,electron-main,browser,node}/**/{*[sS]ervice*,*[cC]ontroller*}.ts
{src,extensions}/**/{common,browser,electron-sandbox}/**/*{[cC]ontribution,[sS]ervice}.ts
src/vs/{base,platform,workbench}/**/{test,browser}/**/*{[mM]odel,[cC]ontroller}*.ts
extensions/**/{browser,common,node}/{**/*[sS]ervice*,**/*[pP]rovider*}.ts
\end{minted}
\caption{Patterns used to benchmark Glob matchers}
\label{lst:bench-patterns}
\end{listing}

\begin{figure}
\centering
\includesvg[width=\textwidth]{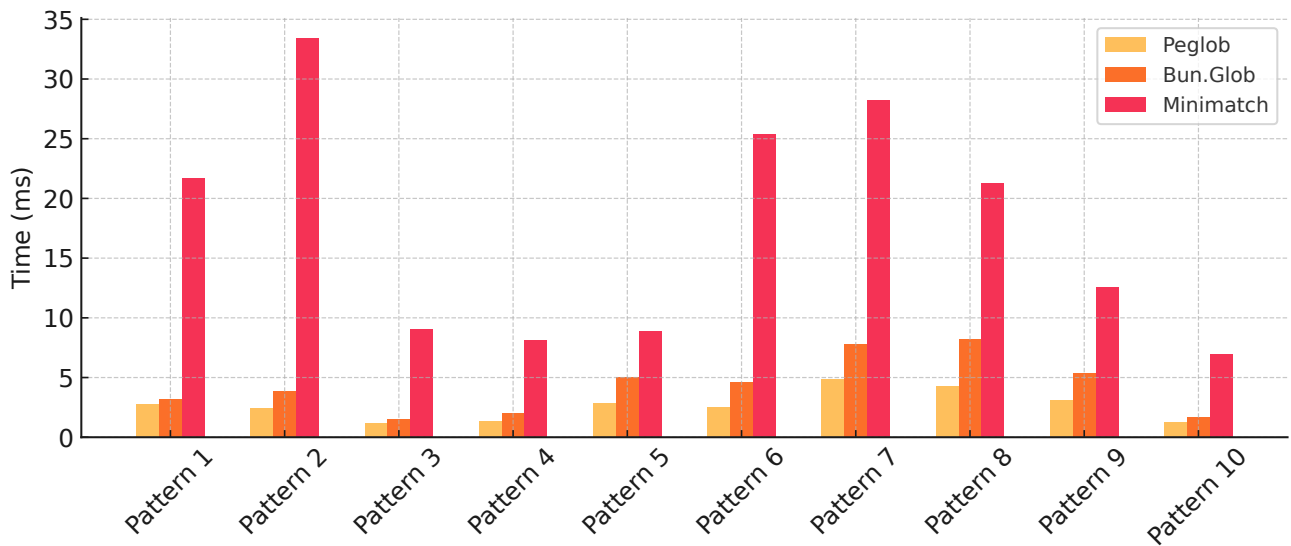}
\caption{Glob performance against VSCode dataset (lower is better)}
\label{fig:glob-performance}
\end{figure}

For performance testing, we used VSCode 1.97.0's source code repository, comprising 7,894 file paths. We designed 10 challenging patterns (shown in Listing~\ref{lst:bench-patterns}) to test real-world performance scenarios. We believe real-world performance is the most valuable metric as it involves a mix of matching and non-matching paths. We deliberately created high-difficulty test cases because performance differences in glob matching typically emerge from stars, globstars, and braced conditions, so our examples extensively combine these syntaxes while remaining practical.

We compared Peglob with Bun.Glob\footnote{Using Bun v1.2.4-canary.13+6aa62fe4b which includes the \texttt{skip\_brace} bugfix after Bun 1.2.3 that ported the Rust crate fast-glob.} and Minimatch\footnote{Using Minimatch 10.0.1, run in Bun environment for consistent comparison.}, excluding other Lua glob libraries that couldn't correctly handle most of our test patterns. We chose Bun.Glob partly because it met our correctness requirements, thanks in part to our feedback to the Bun team.\footnote{Github Issue: Inconsistent Glob Pattern Matching Results in Bun 1.2.2 and 1.2.3-canary. \url{https://github.com/oven-sh/bun/issues/17512}} Minimatch, a JavaScript library that excels in correctness, represents a typical approach of converting globs to equivalent regular expressions. Importantly, our benchmark measured only the final matching speed, not including time spent converting globs to LPeg objects or regular expressions.

The results demonstrate that our optimized Peglob implementation consistently outperformed both competitors across all test patterns. Peglob is 14\% to 92\% faster than Bun.Glob and 3 to 14 times as fast as Minimatch. The performance gap between Peglob and its competitors widens as pattern complexity increases, with Bun.Glob struggling with specific patterns (e.g. Pattern 7 and Pattern 8) and Minimatch showing high computational overhead across all scenarios.

\begin{figure}
\centering
\includesvg[width=0.8\textwidth]{glob-perf1.svg}
\includesvg[width=0.8\textwidth]{glob-perf2.svg}
\includesvg[width=0.8\textwidth]{glob-perf3.svg}
\includesvg[width=0.8\textwidth]{glob-perf4.svg}
\includesvg[width=0.8\textwidth]{glob-perf5.svg}
\caption{Glob performance in edge cases (lower is better). Each row shows one test group with matching scenarios (left: patterns expected to match; right: patterns expected not to match, potentially triggering backtracking). Note: Bun.Glob results in the last two rows (testing braced conditions) stop at $n=10$, reflecting its documented limit of 10 nested brace levels.\citep{bun-glob}}
\label{fig:glob-performance2}
\end{figure}

Beyond real-world scenarios, we also evaluated Peglob's performance in extreme cases prone to backtracking. We designed five test groups to measure processing speed when gradually increasing variable-length elements for large inputs, inspired by Cox's work \citep{Cox2017}. Each test group contains two scenarios: one where the pattern is expected to match the input (requiring less backtracking), and another where the pattern is expected not to match (potentially triggering extensive backtracking in unoptimized implementations).

The results in Figure~\ref{fig:glob-performance2} show Minimatch lacks optimizations for these situations, demonstrating what performance would look like without relevant optimizations. Both Peglob and Bun.Glob performed excellently, avoiding excessive backtracking. For the last test group, specifically designed to test braced condition optimization effectiveness, Peglob's performance remained remarkably stable. Without our braced condition optimization, Peglob would exhibit exponential time complexity due to numerous branches after brace expansion.

These consistent results across correctness and performance metrics demonstrate that our formal grammar specification and architectural decisions successfully balance robustness with efficiency, while our optimization strategies deliver significant performance improvements without compromising reliability.

\section{Conclusion}
\label{conclusion}

This paper has explored advanced techniques for leveraging LPeg's capabilities through two case studies: a high-performance JSON parser and a sophisticated Glob-to-LPeg pattern converter.

Our JSON parser analysis revealed that substitution capture significantly improves efficiency by reducing memory allocation overhead, particularly for string-heavy JSON documents. The table construction optimization demonstrated measurable performance improvements for object-intensive documents by pre-allocating hash space and optimizing C-to-Lua transitions. These optimizations enabled our parser to outperform other pure Lua implementations while highlighting areas where native C implementations still maintain advantages.

The Glob-to-LPeg converter case study demonstrated how a comprehensive approach to pattern matching can lead to effective results. By establishing a formal grammar specification, implementing Cox's flattened search strategy, optimizing braced conditions, and carefully handling edge cases, we developed a solution that consistently performed well in both real-world and stress-test scenarios. Our evaluation against the VSCode repository showed Peglob performing 14\% to 92\% faster than Bun.Glob and 3 to 14 times faster than Minimatch across diverse pattern matching challenges. The architectural decision to separate in-segment pattern matching from segment-crossing behaviors proved particularly effective, enabling targeted optimizations while maintaining pattern matching correctness.

These case studies illustrate that PEG-based parsers can be substantially optimized through technique selection and grammar construction without modifying the underlying library. They also demonstrate that systematic analysis of pattern matching problems can lead to implementations that excel in both performance and maintainability.

Future research could explore enhancing LPeg's memory management efficiency and extending these optimization techniques to other complex parsing domains. Our work provides practical insights for implementing efficient parsers while maintaining code clarity and reliability, contributing to the advancement of text processing capabilities in the Lua ecosystem.

\section{Acknowledgement}
\label{acknowledgement}

We thank Professor Roberto Ierusalimschy and all contributors to the LPeg library for their work in creating this PEG implementation that enabled our research. We are grateful to Dylan Conway of the Bun Team, whose prompt resolution of issues in Bun.Glob enabled us to successfully present a comprehensive benchmark for Glob matchers with all tests accurately executed.

We extend special thanks to Justin M. Keyes and Christian Clason for reviewing and merging the Peglob implementation into Neovim\footnote{https://github.com/neovim/neovim/pull/33605}. The feedback from the Neovim community was invaluable in making Peglob more robust and production-ready.

We also appreciate the reviewers for their insightful and constructive feedback, which significantly enhanced the quality of this paper.

Lastly, we are grateful to the readers of this work. Your interest and engagement give our efforts purpose. We hope this paper serves as a valuable resource and source of inspiration in your exploration of LPeg and PEG, and we invite you to join us in advancing the growth and vitality of the LPeg community.

\section*{Declaration of Generative AI in Scientific Writing}

During the preparation of this work the author used Anthropic Claude in order to translate the manuscript into English and improve the writing. After using this tool/service, the author reviewed and edited the content as needed and takes full responsibility for the content of the published article.

\bibliographystyle{elsarticle-num}
\bibliography{refs}

\end{document}